\documentclass[aps,prb,twocolumn,showpacs,superscriptaddress,citeautoscript,reprint]{revtex4-1}

\usepackage{bm}
\usepackage{graphicx}
\usepackage{amsmath}
\usepackage{amsfonts}

\begin{document}

\title[Enhancing thermoelectric properties of graphene quantum rings]%
{Enhancing thermoelectric properties of graphene quantum rings}

\author{M. Saiz-Bret\'{i}n}

\affiliation{GISC, Departamento de F\'{\i}sica de Materiales, Universidad Complutense, E-28040 Madrid, Spain}

\author{A. V. Malyshev}

\affiliation{GISC, Departamento de F\'{\i}sica de Materiales, Universidad Complutense, E-28040 Madrid, Spain}

\affiliation{Ioffe Physical-Technical Institute, 26 Politechnicheskaya Street, 194021 St. Petersburg, Russia}

\author{P. A. Orellana}

\affiliation{Departamento de F\'{\i}sica, Universidad T\'{e}cnica Federico Santa Mar\'{\i}a, Casilla 110 V, Valpara\'{\i}so, Chile}

\author{F. Dom\'{\i}nguez-Adame}

\affiliation{GISC, Departamento de F\'{\i}sica de Materiales, Universidad Complutense, E-28040 Madrid, Spain}

\affiliation{Department of Physics, University of Warwick, Coventry, CV4 7AL, 
United Kingdom}


\begin{abstract}

We study the thermoelectric properties of rectangular graphene rings connected symmetrically or asymmetrically to the leads. A side-gate voltage applied across the ring allows for the precise control of the electric current flowing through the system. The transmission coefficient of the rings manifest Breit-Wigner line-shapes and/or Fano line-shapes, depending on the connection configuration, the width of nanoribbons forming the ring and the side-gate voltage. We find that the thermopower and the figure of merit are greatly enhanced when the chemical potential is tuned close to resonances. Such enhancement is even more pronounced in the vicinity of Fano like anti-resonances which can be induced by a side-gate voltage independently of the geometry. This opens a possibility to use the proposed device as a tunable thermoelectric generator.

\end{abstract}

\maketitle

\section{Introduction}

One of the main goals in thermoelectric research is to find materials and devices with a high figure of merit $ZT=S^2\sigma T / \kappa$, which reflects the thermoelectric efficiency of the system~\cite{Goldsmid10}. Here $S$ is the thermopower (Seebeck coefficient), and $\sigma$ and $\kappa$ are the electric and thermal conductances at a given temperature $T$, respectively. Since the advent of nanotechnology, many discoveries have demonstrated that nanometer-sized objects exhibit physical properties not shared by bulk materials~\cite{Bhusan03}. In particular, theoretical predictions~\cite{Hicks93,Khitun00,Balandin03} and experiments~\cite{Venkata01,Harman02,Hochbaum08,Boukai08} pointed out that thermoelectric properties at the nanoscale are strongly enhanced. Efficient thermoelectric bulk materials display values of $ZT$ lower than unity. However, Venkatasubramanian \emph{et al.} reported $ZT=2.4$ in thin-film superlattices at room temperature~\cite{Venkata01}. Similarly, Harman \emph{et al.} studied a quantum dot superlattice and found  $ZT=1.6$~\cite{Harman02}. Therefore, nanostructures pave a possible way to achieve large $ZT$ and consequently more efficient thermoelectric devices as refrigerators and generators~\cite{Koumoto13}.

The enhancement of the figure of merit in nanodevices can be caused by different mechanisms. Both electrons and phonons contribute to the thermal contribution ($\kappa=\kappa_e + \kappa_\mathrm{ph}$) and lower values of any of them will yield higher values of $ZT$. Kithun \emph{et al.} reported high values of the figure of merit in a quantum-dot superlattice due to strong phonon scattering.~\cite{Khitun00} Chang and Nikoli\'{c} have demonstrated that nanopore arrays in graphene nanoribbons can block phonons while retaining edge electron currents, yielding $ZT \approx 5$.\cite{Chang12} An alternative route to achieve larger $ZT$ relies on engineering the electronic properties of the nanodevices. In this regard, Murphy \emph{et al.} have shown that the increase of the density of states at the Fermi level in molecular junctions may result in an increased thermopower $S$~\cite{Murphy08}. Sharapov and Varlamov have predicted a considerable enhancement of the thermoelectric power in gapped graphene in the vicinity of the gap edges due to some peculiarities of quasiparticle scattering from impurities.\cite{Sharapov12} On the other hand, quantum interference effects can also play a role in the observed thermoelectric properties of single-molecule heterojunctions~\cite{Bergfieldand09} and zero-dimensional systems~\cite{Karlstrom11}. In particular, quantum effects giving rise to Fano resonances in the transmission probability were predicted to have an impact on the thermopower magnitude and the thermoelectric efficiency of quantum dot systems~\cite{Nakanishi07,Trocha12,Gomez-Silva12}, single-molecule devices~\cite{Finch09} and nanoscale junctions~\cite {Garcia-Suarez13}.

Graphene is a material with a combination of many remarkable properties~\cite{Castro-Neto09} and has recently received attention as a potential thermoelectric material too. In particular, thermoelectric phenomena in graphene nanoribbons have been widely studied~\cite{Divari10,Ouyang09,Rosales13,Sevincli10,Mazzamuto11}. In this context, Mazzamuto \emph{et al.} have analyzed thermal and electrical properties of perfect graphene nanoribbons as a function of their width and edge orientation~\cite{Mazzamuto11}. They found that a structure with armchair and zigzag sections of different widths presents lower phonon thermal conductance while retaining high electronic conductance and thermopower, yielding $ZT$ that exceeds unity at room temperature. Another approach to achieve higher thermoelectric efficiencies in graphene is to exploit quantum coherence effects such as the aforementioned Fano effect. Its large electron mobility and long coherence lengths make graphene an ideal candidate for the usage of these quantum phenomena.

Recently, we proposed a novel design of a quantum interference device based on a hexagonal graphene ring~\cite{Munarriz11,Munarriz12}, in which all edges are of the same type to reduce scattering at bends~\cite{Fertig10}. In this paper we study the thermoelectric properties of rectangular graphene quantum rings because they have both type of edges in the same device, aiming to reduce phonon thermal conductance, as suggested in Ref.~\onlinecite{Mazzamuto11}.  We study two configurations, namely, symmetrically and asymmetrically connected square nanorings (see Fig.~\ref{fig1}) and conclude that the latter displays higher thermoelectric efficiency due to the occurrence of Fano anti-resonances in the transmission. We also show how the thermoelectric coefficients can be controlled by a side-gate voltage.

\section{Model and formalism}

\begin{figure}[ht]
\begin{center}
\includegraphics[width=\linewidth]{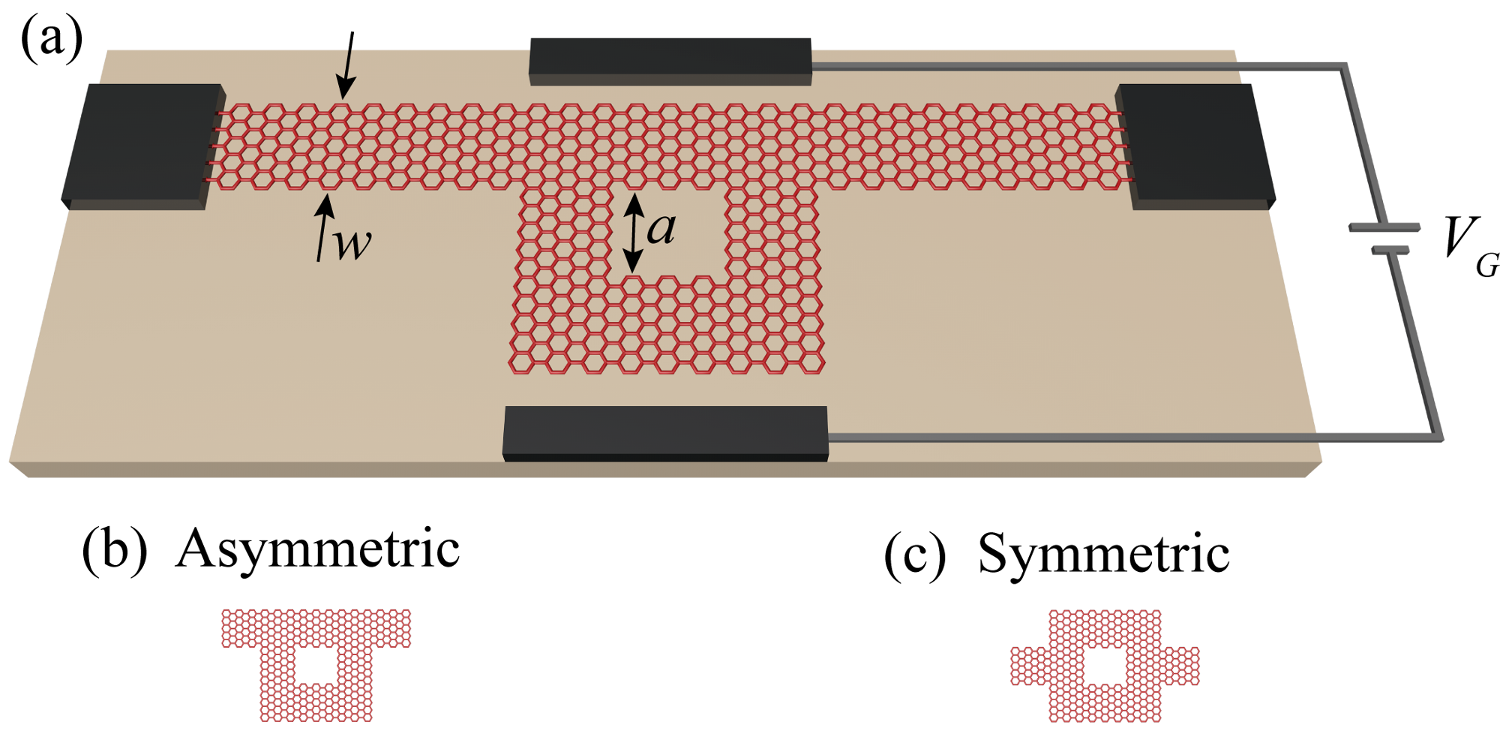}
\end{center}
\caption{ (Color online) (a) Schematic diagram of the graphene nanoring connected to two armchair nanoribbons. A side-gate voltage can be applied across the ring. The connection can be (b)~asymmetric or (c)~symmetric.}
\label{fig1}
\end{figure}

The system under consideration consists of a square graphene nanoring connected symmetrically or asymmetrically to two leads, as shown in Fig.~\ref{fig1}. The width of all nanoribbons, both in the leads and in the ring, is $w$ and the dimensions of the inner hole are $a\times a$. For concreteness, we restrict ourselves to the case $w=a$.

We assume that the leads are two semi-infinite armchair graphene nanoribbons with $N \neq 3n-1$, $N$ being the number of hexagons across the nanoribbon and $n$ a positive integer. In this case the band structure has a width-dependent gap and the corresponding dispersion relation in the vicinity of the gap is parabolic. This energy spectrum is typical for conventional semiconductors and we showed that it presents more robust and promising transmission patterns for transport control and applications~\cite{Munarriz11}. 

The devices are modeled using a nearest neighbor tight-binding Hamiltonian
\begin{equation}
\mathcal{H}=\sum_{i}\epsilon_i|i \rangle\langle i| -t\sum_{\langle i,j\rangle}|i\rangle\langle j|\ ,
\end{equation}
where the hopping parameter is set to $t=2.8\,$eV and the site energy $\epsilon_i$ can vary from one atom to another due to the presence of a side-gate voltage drop $V_G$ across the ring. The profile of the electric field can be calculated by solving the Poisson and Schr\"{o}dinger equations self-consistently. However, for simplicity, we assume a simplified side-gate potential profile: it is linear in the transverse direction (perpendicular to the electric current) while in the longitudinal direction it is constant within the nanoring area and decays exponentially toward the two leads. This voltage allows us to control the current in the device, as thoroughly discussed in Refs.~\onlinecite{Munarriz11,Munarriz12}.

Assuming that electron-phonon scattering in our samples is reduced, we consider electrons in the fully coherent regime transferring ballistically through the system. The quantum transmission boundary method~\cite{Lent90,Ting92} combined with the effective transfer matrix method~\cite{Schelter10} are used to compute wave functions and the transmission coefficient $\tau(E)$ as a function of energy (see Ref.~\onlinecite{Munarriz14} for further details).  Once $\tau(E)$ is known, the thermopower $S$, the electric conductance $\sigma$ and the electric thermal conductance $\kappa_e$ (we neglect the phonon contribution) can be obtained from the following expressions~\cite{Gomez-Silva12}
\begin{subequations}
\begin{eqnarray}
S&=&-\frac{1}{eT}\,\frac{K_1}{K_0}\ ,\\
\sigma &=& e^2 K_0\ ,\\
\kappa_e &=& \frac{1}{T}\,\left(K_2-\frac{K_1^2}{K_0}\right)\ ,
\end{eqnarray}
where we have introduced the notation
\begin{equation}
K_n=\frac{2}{h}\int_{-\infty}^{\infty}\left(-\frac{\partial f}{\partial E}\right)(E-\mu)^n\tau(E)\,dE\ .
\end{equation}
\label{thermoelectric}
\end{subequations}
In this equation $\mu$ is the chemical potential of the graphene leads, $f(E)=\big\{\exp\big[(E-\mu)/k_BT\big]+1\big\}^{-1}$ is the Fermi distribution function and $k_B$ is the Boltzmann constant. For concreteness we take $T=4\,$K hereafter.

\section{Results}

\begin{figure}[ht]
\begin{center}
\includegraphics[width=\linewidth]{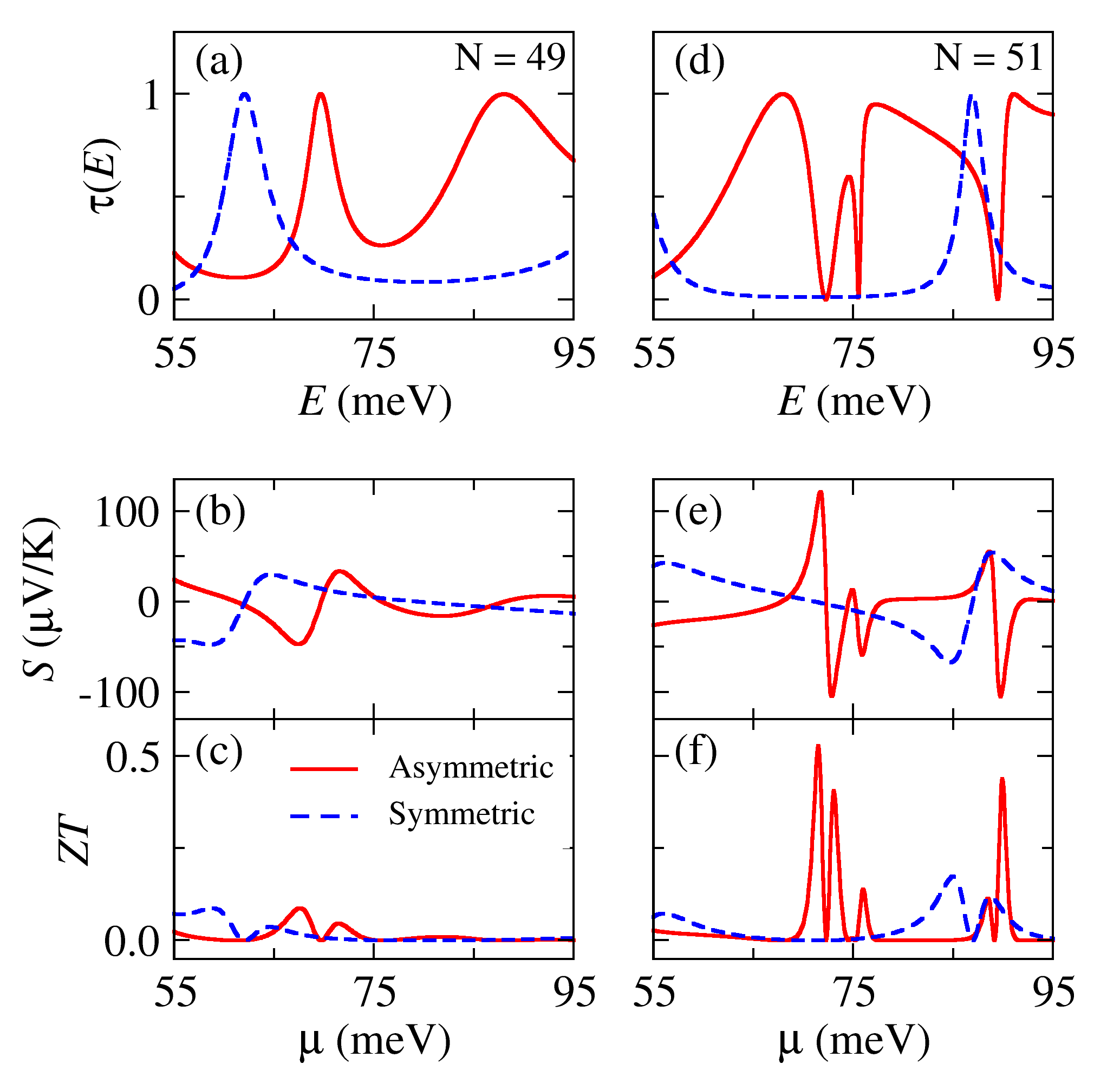}
\end{center}
\caption{(Color online) Transmission and thermoelectric coefficients in the absence of the side-gate voltage ($V_G=0$) for symmetric (dashed lines) and asymmetric (solid lines) rings. (a)~Transmission coefficient, (b)~Seebeck coefficient and (c)~figure of merit for a nanoribbon width of $N=49$. (d)~Transmission coefficient, (e)~Seebeck coefficient and (f)~figure of merit for a nanoribbon width of $N=51$.}
\label{fig2}
\end{figure}

First, we compare the transmission properties of symmetric and asymmetric rings in the absence of side-gate voltage ($V_G=0$). The focus is set at the one-mode regime in which interference related effects are not smoothed out due to the superposition of several modes. We have numerically found that the transmission patterns can be grouped into two categories, depending on the value of $N$. If $N=3n-2$ the transmission coefficient displays resonant peaks whose shape is Lorentzian close to the resonance energy for both configurations (Breit-Wigner line-shapes). A typical example is shown in Fig.~\ref{fig2}(a) corresponding to $w=12.0\,$nm, i.e., $N=49$, for both symmetric (dashed line) and  asymmetric (solid line) rings. When $N=3n$ the transmission coefficient strongly depends on the symmetry of the ring. As shown in Fig.~\ref{fig2}(d) for $w=12.5\,$nm, i.e., $N=51$, the transmission coefficient for symmetrically connected rings only presents Breit-Wigner line-shapes (dashed line). On the contrary, if the ring is connected asymmetrically, the transmission coefficient shows Fano line-shapes (solid line). When the nanoribbon width is increased, the one-mode energy region shrinks, but the transmission features remain qualitatively unchanged.

According to Ref.~\onlinecite{Garcia-Suarez13}, Fano and Breit-Wigner like resonances are expected to enhance the thermopower and figure of merit of nanoscale junctions when the chemical potential crosses these features in the transmission. We will see that this statement is also correct for graphene nanorings. When the chemical potential is close to a Breit-Wigner like resonance, such as those in Fig.~\ref{fig2}(a) at about $62\,$meV and $70\,$meV or in Fig.~\ref{fig2}(d) at $87\,$meV, there is an enhancement in the thermopower $S$ and the figure of merit $ZT$. Notice that narrower resonances give a larger increment of the thermoelectric coefficients. It is worth mentioning that Mahan and Sofo also showed that in general the best figure of merit is achieved for a narrow resonance located at the Fermi energy.\cite{Mahan96} However, the figure of merit remains rather small ($ZT \leq 0.15$). The enhancement due to Fano like anti-resonances is remarkable compared to the case of Breit-Wigner like resonances. Figures~\ref{fig2}(e) and \ref{fig2}(f) shows that the thermopower and figure of merit reach a maximum value close to $|S|\simeq 100\,$ $\mu$V/K and $ZT\simeq 0.5$ when the Fano like anti-resonances at $72\,$meV and $90\,$meV cross the chemical potential. We now argue why asymmetric resonances are preferred to obtain better thermoelectric devices. The enhancement of $ZT$ due to Fano like anti-resonances compared to Breit-Wigner like resonances can be understood from Eqs.~(\ref{thermoelectric}). The figure of merit is proportional to the thermopower $S$, namely it is proportional to $K_1$. The product $\left(-\partial f/\partial E\right)(E-\mu)$ appearing in the expression of $K_1$ is an odd function with respect to the variable $E-\mu$. Therefore, the larger the asymmetry of $\tau(E)$ around the chemical potential, the higher the thermoelectric response of the nanosystem.

\begin{figure}[ht]
\begin{center}
\includegraphics[width=0.8\linewidth]{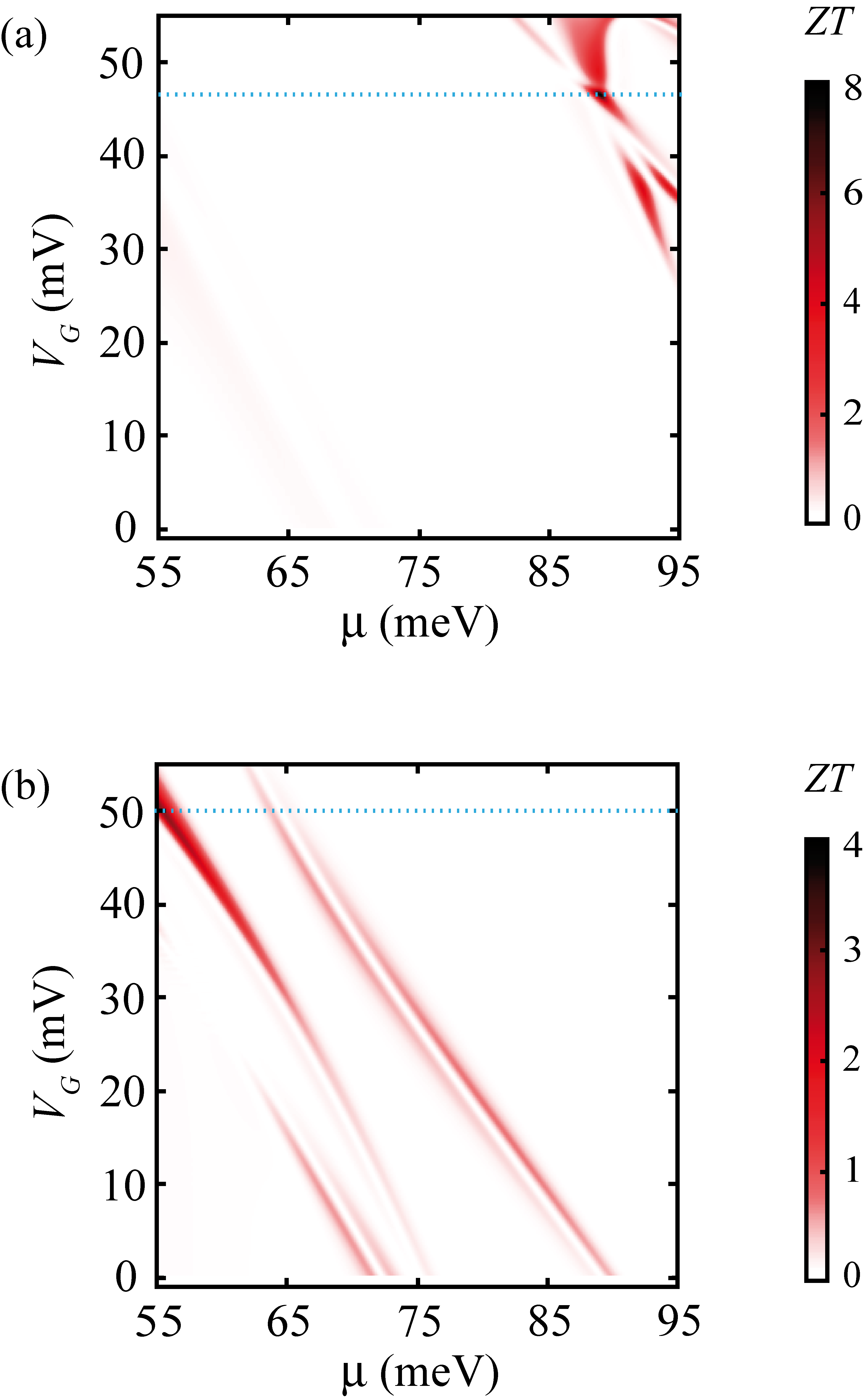}
\end{center}
\caption{(Color online) Figure of merit $ZT$ as a function of the chemical potential $\mu$ and the side-gate voltage $V_G$ for the asymmetric rings with (a)~$N=49$ and (b)~$N=51$. The dotted lines mark the cross sections shown in Fig.~\ref{fig4}. }
\label{fig3}
\end{figure}

\begin{figure}[ht]
\begin{center}
\includegraphics[width=\linewidth]{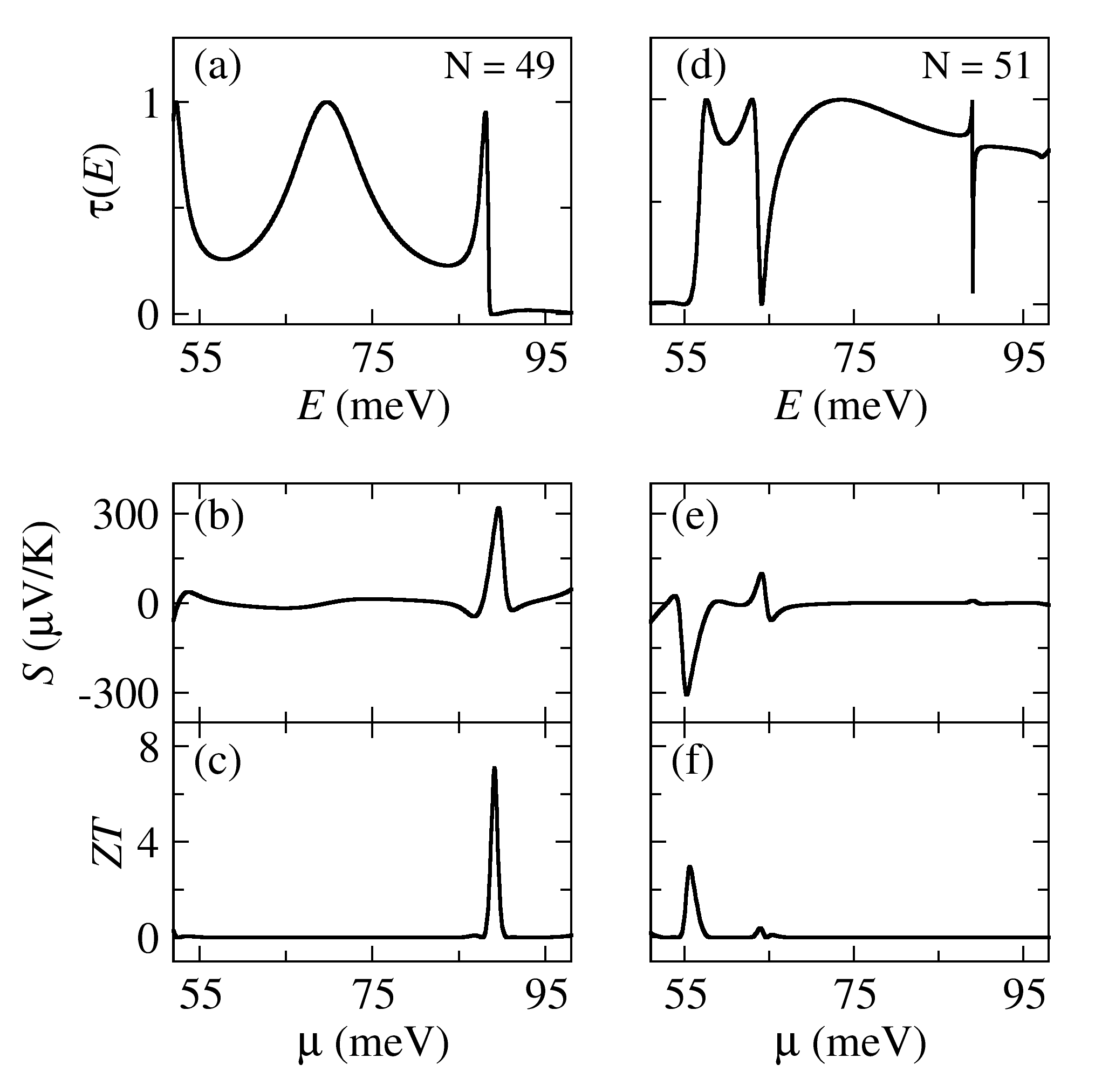}
\end{center}
\caption{(a) Transmission coefficient, (b)~Seebeck coefficient and (c) figure of merit for the cross section marked in Fig.~\ref{fig3}(a) that corresponds to a nanoribbon width of $N=49$ and a side-gate voltage drop $V_G=46\,$mV. (d)~Transmission coefficient, (e)~Seebeck coefficient and (f)~figure of merit for the cross section marked in Fig.~\ref{fig3}(b) that corresponds to a nanoribbon width of $N=51$ and a side-gate voltage drop $V_G=50\,$mV.}
\label{fig4}
\end{figure}

Looking for a stronger increase of the thermoelectric performance, we applied a side-gate voltage $V_G$ as shown schematically in Fig.~\ref{fig1}(a). Figure~\ref{fig3} shows the figure of merit for the asymmetric ring with $N=49$ (upper panel) and $N=51$ (lower panel) as a function of the chemical potential $\mu$ and the side-gate voltage $V_G$. Tuning both $\mu$ and $V_G$, the thermoelectric coefficients can be controlled and, in particular, the figure of merit can be grately enhanced. For the ring with $N=51$, the figure of merit can reach values close to $ZT\simeq 3$ when a side-gate voltage of $V_G=50\,$mV is applied.  In the case of $N=49$, a side-gate voltage of $V_G=46\,$mV leads to a maximum value of $ZT\simeq 7$. Figures~\ref{fig4}(a) and (d) show the transmission coefficients corresponding to the cross sections marked by dotted lines in Fig.~\ref{fig3}. These transmission coefficients display new sharp Fano line-shapes that arise due to $V_G$. The occurrence of these sharp Fano resonances gives rise to the increase of both the thermopower $S$ and the figure of merit $ZT$. Similar results were found for the symmetric rings (not shown here). As expected, the good thermoelectric performance is degraded on increasing temperature. We found that the maximum figure of merit $ZT$ is of the order of unity above $50\,$K.

\begin{figure}[ht]
\begin{center}
\includegraphics[width=\linewidth]{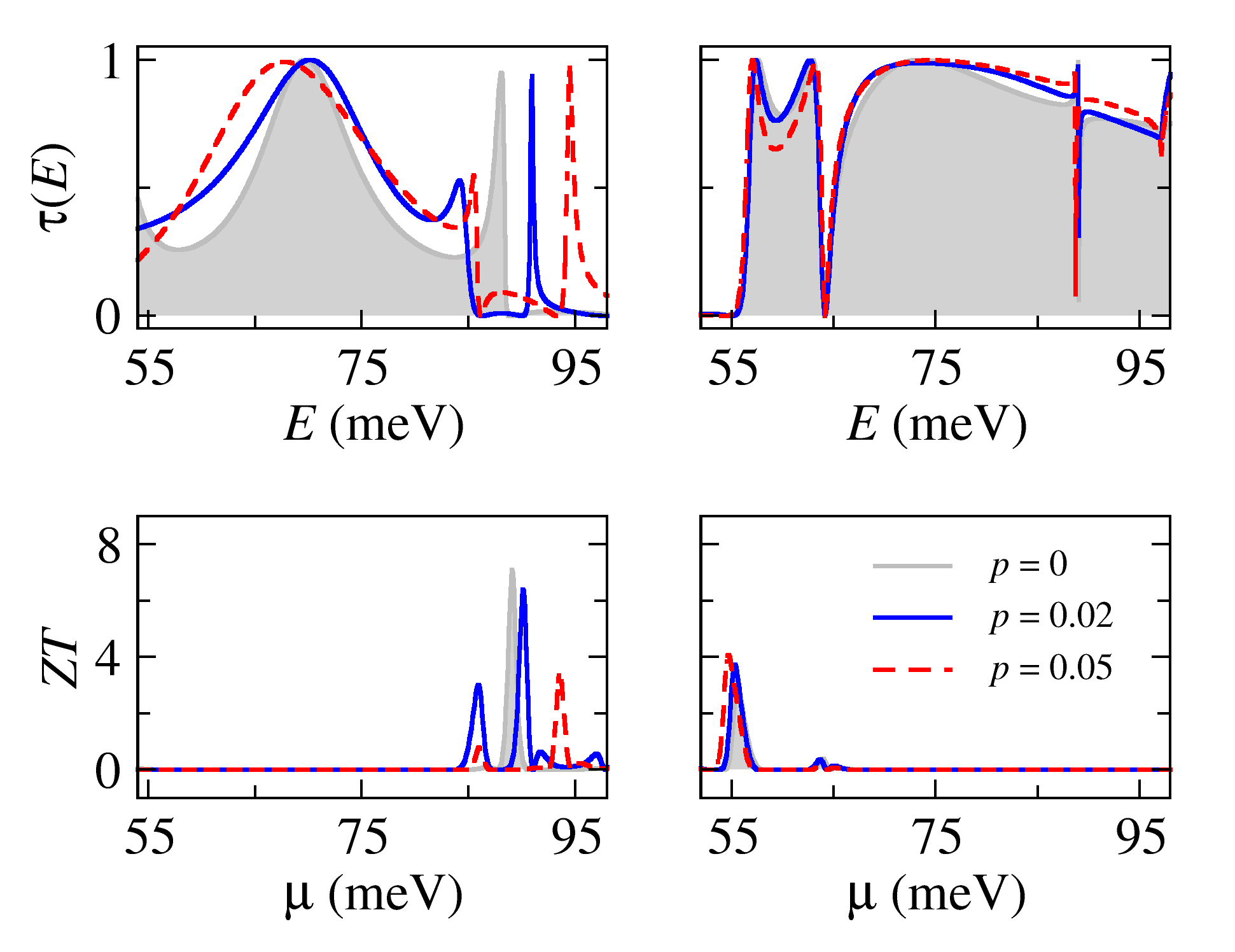}
\end{center}
\caption{(Color online) Upper panels show the transmission coefficient as a function of energy for $N=49$ with $V_G=46\,$mV (left), and $N=51$ with $V_G=50\,$mV (right). Lower panels show the corresponding figure of merit as a function of the chemical potential. Grey solid lines with filling correspond to samples with ideal edges ($p=0$), while red dashed lines and blue solid lines correspond to edge-disordered ones with $p=0.02$ and $p=0.05$, respectively.}
\label{fig5}
\end{figure}

Finally we address the effects of the edge disorder on the thermoelectric properties of the graphene rings. We randomly remove atoms of carbon from the zig-zag edges with some given probability $p$. In the case of armchair edges, pairs of neighbor atoms are removed with the same probability. By removing pairs rather than single atoms we ensure that there are no dangling atoms in the sample, so we do not have to deal with complicated edge reconstruction effects. The transmission coefficient and the corresponding figure of merit calculated for typical realizations of disorder are shown in Fig.~\ref{fig5}. When $N=3n-2$ the figure of merit can be degraded but still remains high and new resonances can appear in the transmission coefficient. This is due to the fact that we were removing atoms from the edges, making the nanoribbons effectively narrower. Because graphene nanoribbons have a width-dependent gap, these narrower areas will act as barriers, causing new resonances to appear. On the contrary, when $N = 3n$ the transmission coefficient and figure of merit are rather insensitive to disorder. For these nanoribbons, removing edge atoms leads to effective widths of $N=3n-1$ in some areas which can be expected to act as less efficient barriers because a nanoribbon of such a width has a gapless spectrum. Consequently, the transmission coefficient is more robust to the moderate disorder for the case $N=3n$. 

We note that the contribution of phonons to the thermal conductance $\kappa_\textrm{ph}$ has been neglected in our calculations. This contribution would lead to a smaller figure of merit because it increases the denominator of $ZT$. However, careful engineering of the type of edges can reduce the effects of phonons, as suggested in Ref.~\onlinecite{Mazzamuto11}. Similar conclusions were recently drawn by Li \emph{et al}, who found that the thermal conductance contributed by phonons is greatly reduced in double-bend graphene nanoribbons.\cite{Li14} The reduction was attributed to the influence of the bends. We assume therefore that in the case of square rings, formally formed by four double-bend graphene nanoribbons, the thermal conductance contribution from phonons can expected to be low too. Edge disorder is an alternative route to improve the thermoelectric response of graphene nanoribbons since it also reduces the phonon thermal conductance while retaining quasiballistic electron transport~\cite{Haskins11}. In contrast to graphene nanoribbons,\cite{Chang12} we have found that the electronic properties of graphene quantum rings are not markedly deteriorated by edge disorder. Moreover, the improvement of the thermopower close to a Fano anti-resonance is independent of the phonon contribution, so we can conclude that these resonant energies are good candidates for optimizing the thermoelectric performance of the system. Most importantly, our design yields higher values of $ZT$ (up to $8$) compared to previous proposals. For instance, arrays of nanopores in graphene nanoribbons reduces the phonon contribution but the maximum figure of merit is $ZT\simeq 5$.\cite{Haskins11} Double-bend graphene nanoribbons yield even lower values of $ZT\simeq 0.3$.\cite{Li14} This result is similar to what reported above for square graphene rings when $V_G=0$ (see Fig.~\ref{fig2}). However, the application of the side-gate voltage raises the figure of merit to much higher values, not shared by previous proposals (see Fig.~\ref{fig4}).

\section{Conclusions}

In summary, we have studied the thermoelectric properties of symmetrically and asymmetrically connected quantum rings based on graphene. The transmission coefficient of the rings manifests Breit-Wigner line-shapes in the former case and can show Fano line-shapes in the latter (depending on the width of the nanoribbons comprising the rings). While Breit-Wigner line-shapes lead to a moderate thermoelectric response, the occurrence of Fano line-shapes causes a dramatic enhancement of the thermoelectric efficiency of these nanodevices. Therefore the non-gated asymmetric rings seem to be more promising from the point of view of applications. However, even if a non-gated ring does not support Fano anti-resonances (e.g., symmetrically connected rings), the application of a side-gate voltage can introduce asymmetry in the system and induce such features in the transmission spectrum, which consequently leads to an enhancement of the thermoelectric response. We have also shown that the predicted effects are robust under moderate edge disorder. Precise positions of the Fano features can be very highly dependent on the system geometry, fabrication imperfections (disorder) and the gate voltage. However, the enhancement of the thermopower and the figure of merit in the energy windows of small transmission coefficient is a common feature, in particular in the vicinity of a Fano anti-resonance. Therefore the principal possibility to induce such features by electrostatic means seems very promising from the point of view of control and optimization of thermoelectric properties of various nanoscopic devices.

\acknowledgments

Work in Madrid was supported by MINECO (projects MAT2010-17180 and MAT2013-46308). A. V. M. was partially supported by CAPES (grant PVE-A121). P.A.O. acknowledges FONDECYT grant 1140571  DGIP/USM  internal grant 11.11.62 and CONICYT ACT 1204. F.\ D-A. thanks the Theoretical Physics Group of the University of Warwick for the warm hospitality. The authors also thank A. M. Goldsborough for the critical reading of the manuscript.

\end{document}